\documentclass[aps,prl,twocolumn,superscriptaddress,showpacs,floatfix]{revtex4}
\usepackage{amsmath}
\usepackage{amssymb}
\usepackage{amsbsy}
\usepackage{euscript}
\usepackage{graphicx}
\usepackage[usenames]{color}

\begin{document}
\title{Estimates of rates for dissociative recombination of NO$_2^+$ $+$ $e^-$ via various mechanisms}

\author{Daniel J.~Haxton}
\affiliation{Lawrence Berkeley National Laboratory, Chemical Sciences, Berkeley, CA 94720, USA}

\author{Chris H. Greene}
\affiliation{Department of Physics, Purdue University, West Lafayette, IN 47907, USA}

\begin{abstract}

We estimate rates for the dissociative recombination (DR) of NO$_2^+$ $+$ $e^-$.  Although accurate excited state potential energy curves for the excited states of the neutral are not available, we estimate that the 1 $^2\Phi_g$ and the 1 $^2\Pi_g$ states of the neutral may intersect the ground state cation potential energy surface near its equilibrium geometry.  Using fixed nuclei scattering calculations we estimate the rate for direct DR via these states and find it to be significant.  We also perform approximate calculations of DR triggered by the indirect mechanism, which suggest that the indirect DR rate for NO$_2$ is insignificant compared to the direct rate. 

\end{abstract}

\maketitle

\section{Introduction}

 NO$_2$ is a fundamental molecule in physical chemistry that is present in appreciable quantities in the Earth's atomsphere.  It has remained one of very few simple triatomics for which the dissociative recombination process has received virtually no attention from either experiment or theory.    The present work is intended to provide a pilot study of this molecule, to build on other recent progress in understanding triatomic and polyatomic DR~\cite{OrelKokoouline}.  Vital information about NO$_2$ has been gleaned from extensive pioneering studies by 
Grant and coworkers~\cite{grant1990, grant1994, grant1995}.
These publications have studied  in depth the Rydberg spectroscopy of low lying vibrational states and found that both predissociation and autoionization processes contribute to the lifetime of these states, in varying ratios depending upon the vibrational state in question.  One key finding in particular~\cite{grant1995} is 
that bending excitation enhances predissociation.  Moreover, these studies provide evidence that the Born-Oppenheimer potential energy surfaces of certain valence electronic states of the neutral cross that of the ground state of the cation near the Frank-Condon region of the latter.  

The characteristics of molecular Rydberg states are pertinent to the process of dissociative recombination (DR)~\cite{orelbook, larssonreview, kokooulineCPL2011} AB$^+$ + $e^-$ $\rightarrow$ A + B.   Predissociation of Rydberg states by valence states indicates that the direct mechanism~\cite{omalley, omalleytaylor} can contribute to DR; the direct mechanism is competitive only when some excited Born-Oppenheimer electronic states of the neutral are isoenergetic with the ground  state of the cation within the Franck-Condon region of the latter.  Such crossings of potential energy surfaces are responsible both for predissociation of Rydberg states and dissociative recombination (DR) via the direct mechanism.  

The observation of vibrational autoionization in NO$_2$ indicates that the indirect mechanism~\cite{indirect} may also be important.  Most studies thus far on DR via the indirect mechanism (for example, Refs. \cite{larsonorel, wavepacket, gustisuzor, roman}) have concentrated on molecules with light atomic constituents.  We anticipate that that indirect DR proceeds violently and may lead to the severing of multiple chemical bonds in the target~\cite{h3expt, kokoo}.  NO$_2^+$ is both made of heavy atomic constituents and they are quite strongly bonded, having double bonds between the nitrogen and each oxygen, and is therefore a test case for this picture.

Recently considerable theoretical progress has been made in understanding the indirect mechanism and accurately calculating rates for it~\cite{roman, haxtonheh}.  We apply recent theories~\cite{hamilton} to DR of NO$_2^+$ in reduced dimensionality in an effort to quantify any possible contribution of the indirect mechanism to the DR rate.  Grant's studies show that vibrational autoionization is significant for low lying vibrational states of the cation;
%
 thus the implication of Grant's experiments to the question of DR is unclear.

\section{Structure calculations for the direct mechanism}

\begin{figure}
\begin{center}
\begin{tabular}{c}
\resizebox{0.8\columnwidth}{!}{\includegraphics*[0.8in, 0.8in][5.8in,4.0in]{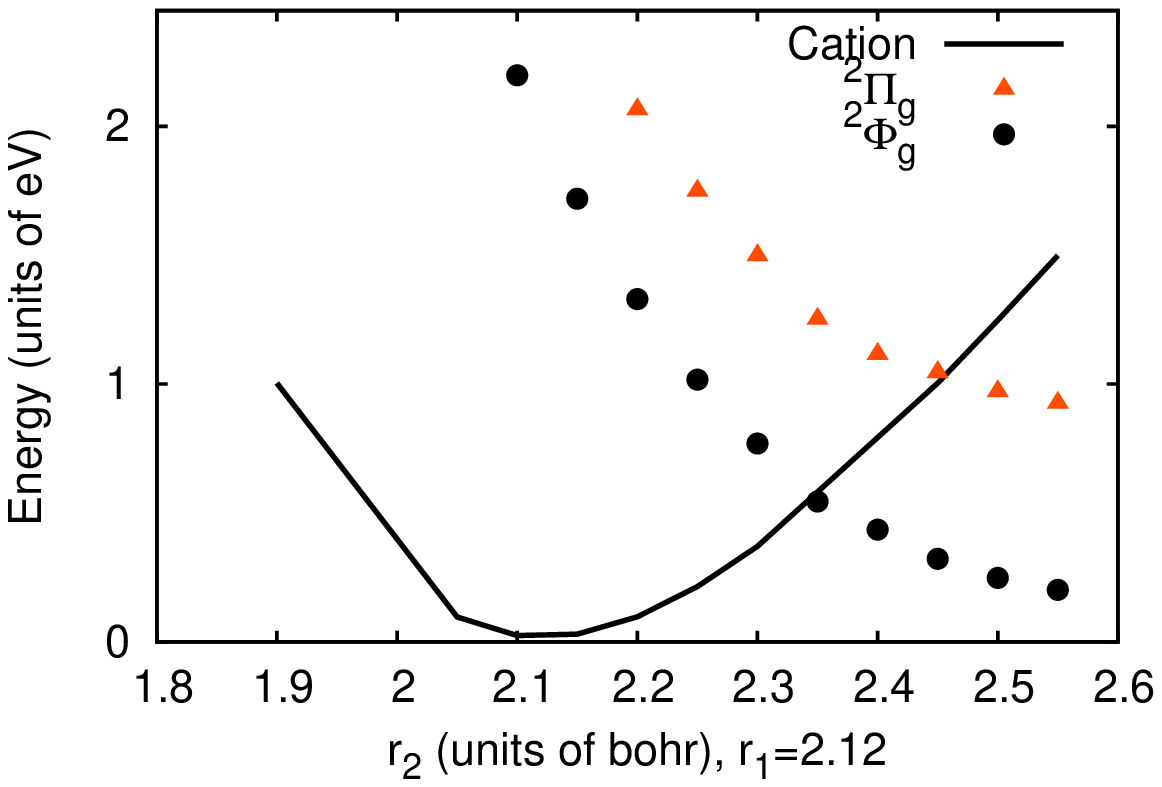}} \\
\resizebox{0.8\columnwidth}{!}{\includegraphics*[0.8in, 0.8in][5.8in,4.0in]{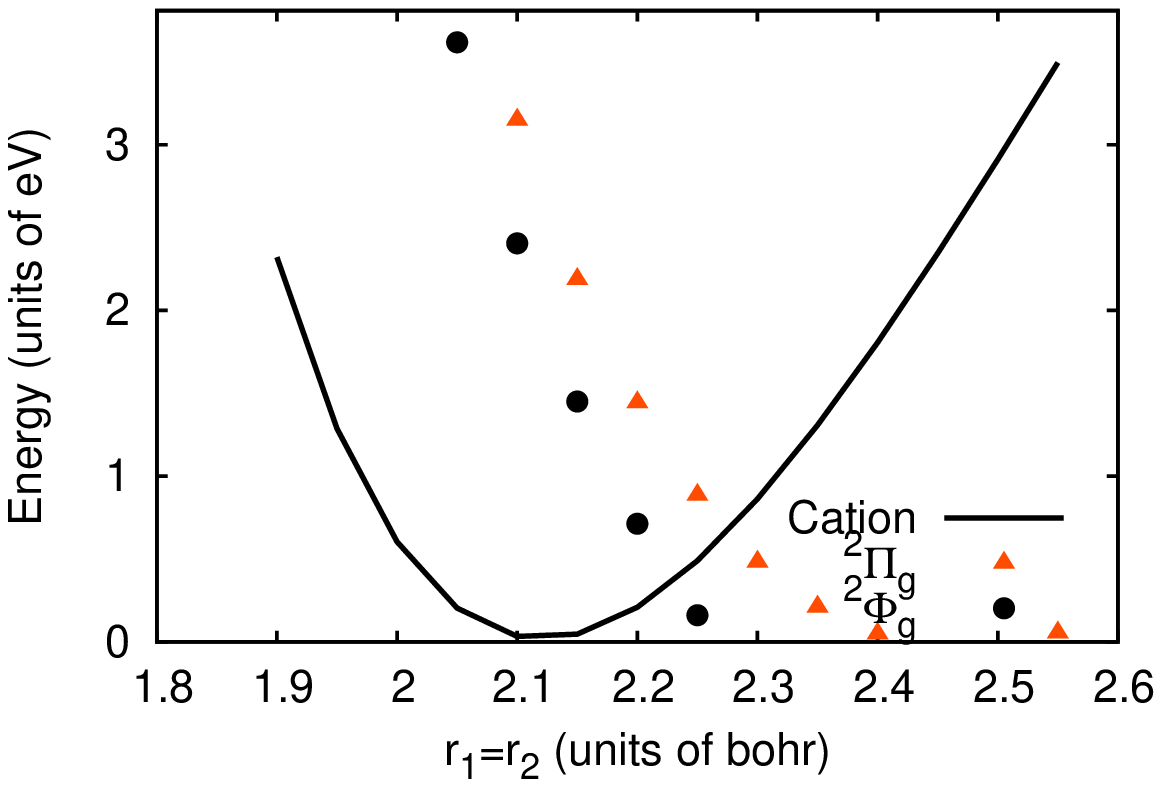}} \\
\resizebox{0.8\columnwidth}{!}{\includegraphics*[0.8in, 0.8in][5.8in,4.0in]{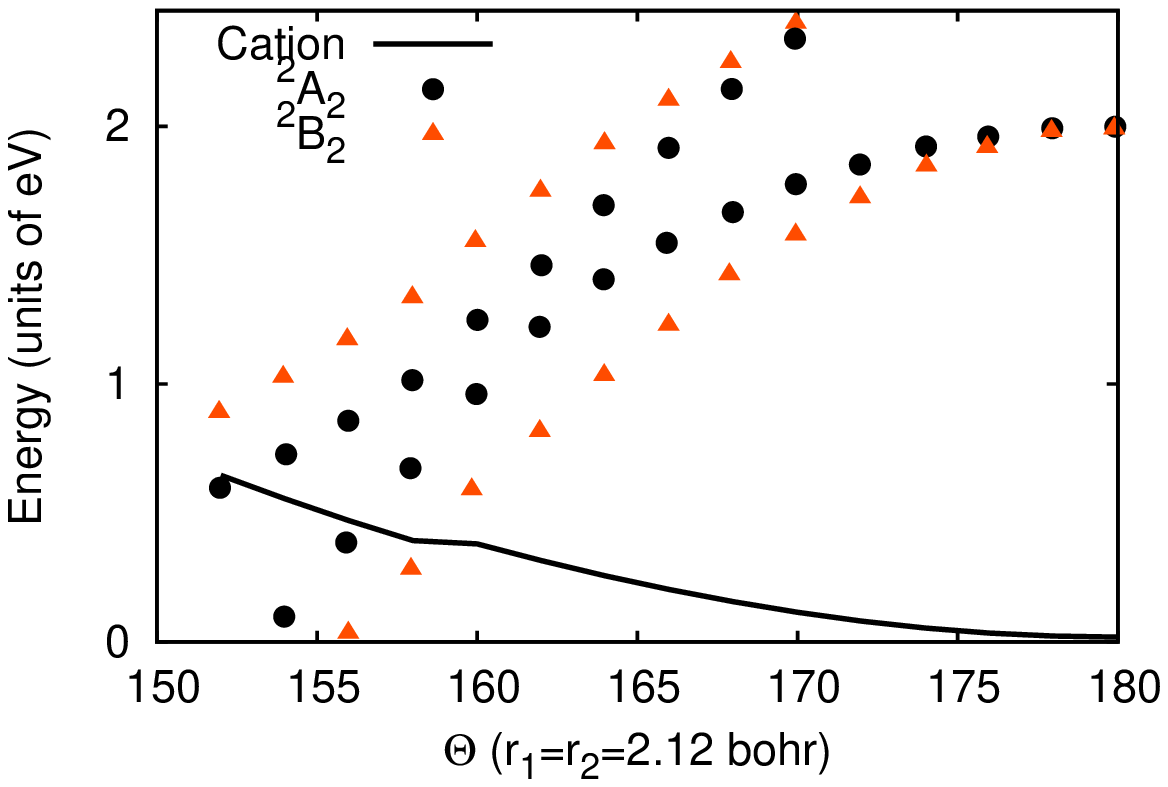}}
\end{tabular}
\end{center}
\caption{(Color online) Structure calculations for the direct mechanism as described in text.~\label{struct}}
\end{figure}

Several authors have looked at the high lying electronic states of NO$_2$, although none has explicitly addressed the difficult question of the energy of the high lying neutral excited states relative to that of the cation.   Katagiri and Kato~\cite{kkato} performed structure calculations on NO$_2$ in the course of a treatment of the photodissociation of this molecule. They analysed the six lowest doublet electronic states of the neutral (lowest at the equilibrium geometry of the neutral).   Kurkal et al.~\cite{kurkal} calculated potential energy surfaces for the lowest two.
Takeshita and Shida~\cite{tshida}
calculated many low-lying electronic states of the cation in the course of a study of photoionization of NO$_2$, and they also analyzed the vibrational structure.

The lowest two states are the ground 1 $^2A_1$ state, which has a bent equilibrium geometry, and the 1 $^2B_1$ state; these both become the X $^2\Pi_u$ state in d$_{\infty H}$ geometry.   The next four states are relevant to the present study.  They are the first and second states of $^2$B$_2$ and $^2$A$_2$ symmetry.  The 1 $^2$A$_2$ and $^2$B$_2$ become the 1 $^2\Phi_g$ state at d$_{\infty H}$ geometry, and the 2 $^2$A$_2$ and $^2$B$_2$ become the 1 $^2\Pi_g$ state at d$_{\infty H}$ geometry.

The ground state configuration of NO$_2^+$ (using labels appropriate to c$_{2V}$ point group symmetry) is (1$a_1^2$ 2$a_1^2$ 1$b2^2$ 3$a_1^2$ 2$b_2^2$ 4$a_1^2$ 3$b_2^2$ 5$a_1^2$ 1$b_1^2$ 4$b_2^2$ 1$a_2^2$), just like the CO$_2$ molecule; the neutral has the 23rd electron in the 6$a_1$ orbital.  The 1 $^2$A$_2$ and $^2$B$_2$ states are described by single excitations from the neutral, 1$a_2$ and 4$b_1$ $\rightarrow$ 6$a_1$, respectively, while the 
2 $^2$A$_2$ and $^2$B$_2$ are described by 4$b_1$ and 1$a_2$  $\rightarrow$ 2$b_1$, respectively.  Relative to the cation, therefore, all of these states are described by double excitations from the NO$_2^+$ + $e^-$ continuum, and we expect them to exist as Feshbach resonances for regions of nuclear geometry in which they are electronically unbound.  As such, they provide good candidates for direct DR.

\section{Structure calculations}

We have performed a set of bound state electronic structure calculations using the COLUMBUS quantum chemistry suite~\cite{colref1,colref2,colref3,colref4,col7} to analyze the crossings of the potential energy surfaces of these states with that of the cation.  These are performed at the configuration-interaction with all singles and doubles (CISD) level of theory and Dunning's aug-cc-pvtz basis set~\cite{dunning}, using a complete active space self-consistent field (CASSCF) reference space with orbitals obtained from a multiconfiguration self-consistent field (MCSCF) calculation.  We include fourteen valence molecular orbitals, leaving out only the highest energy 5$b_2$ (6$\sigma_u$) orbital; we exclude the 1-3$a_1$ and 1-2$b_2$ orbitals from the active space, keeping them doubly occupied, giving CAS(12/9) and (13/9) for the cation and neutral calculations.

For the MCSCF calculation that defines the orbitals we minimize a weighted average of the ground electronic state cation energy and the energies of the first seven states of the neutral (the states listed above plus the $^2\Sigma_g^+$).  In this calculation, the weighted average of the energies of the included
states is minimized with respect to variations of the of orbital and configuration coefficients. The balancing of the (N-1)- and N-electron energies is a difficult problem, but crucial to the question of DR via the direct mechanism.  In order to provide an estimate of the uncertainty in these relative energies, we vary the weighted average; for the seven neutral states we use a factor of 1, and for the cation state we use factors of 3.5, 7, and 14 in the weighted average.

The results are shown in Fig.~\ref{struct}.  We find that the potential energy surfaces of the four high lying electronic states listed above -- 
the 1 and 2 $^2$A$_2$ and $^2$B$_2$ states, which correspond to the $^2\Phi_g$ and $^2\Pi_g$ states at linear geometry -- appear to have crossings with the cation potential energy surfaces within the Franck-Condon region of the latter.   

This bound state structure calculation produces not only eigenvalues approximating the location of metastable states but also discretized continuum states that are unconverged artifacts~\cite{stabilization}.  One spurious root is removed from our plot of bending potential curves in Fig.~\ref{struct}.

\section{Scattering calculations for the direct mechanism}
\label{direct}

Scattering calculations are performed using the polyatomic UK R-matrix code~\cite{ukrmatrix}.  The GTOBAS~\cite{gtobas} code is used to obtain a set of diffuse Gaussians used for the exterior part of the wavefunction.  
The same 14 orbital valence reference space defines the target states of the cation.  These orbitals emerge from an improved 
virtual orbital calculation~\cite{ivo} using neutral Hartree-Fock orbitals.  The Fock operator of which the improved virtual orbitals 
are eigenfunctions is defined as an average of that for the neutral and the cation. 

All four states corresponding to the 1 $^2\Pi_g$ and 1 $^2\Phi_g$ 
states at linear geometry are observed as Feshbach resonances in these calculations.  
 As might be expected, 
because they couple to the $p$-wave continuum,
the components correlating with $^2B_2$ point group take on a larger width than do their $^2A_2$ counterparts, which couple only to partial
waves with $l=2$ and higher in c$_{2V}$ symmetry.  As discussed above, the lower energy 1 $^2\Phi_g$ 
state (the 1 $^2$A$_2$ and 1 $^2$B$_2$ states) becomes 
 isoenergetic with the cation closer to the equilibrium geometry of the neutral than is the 1 $^2\Pi_g$ state.

\section{Estimate of direct DR cross section using the reflection principle \label{direct}}

\begin{figure}
\begin{center}
\resizebox{0.8\columnwidth}{!}{\includegraphics{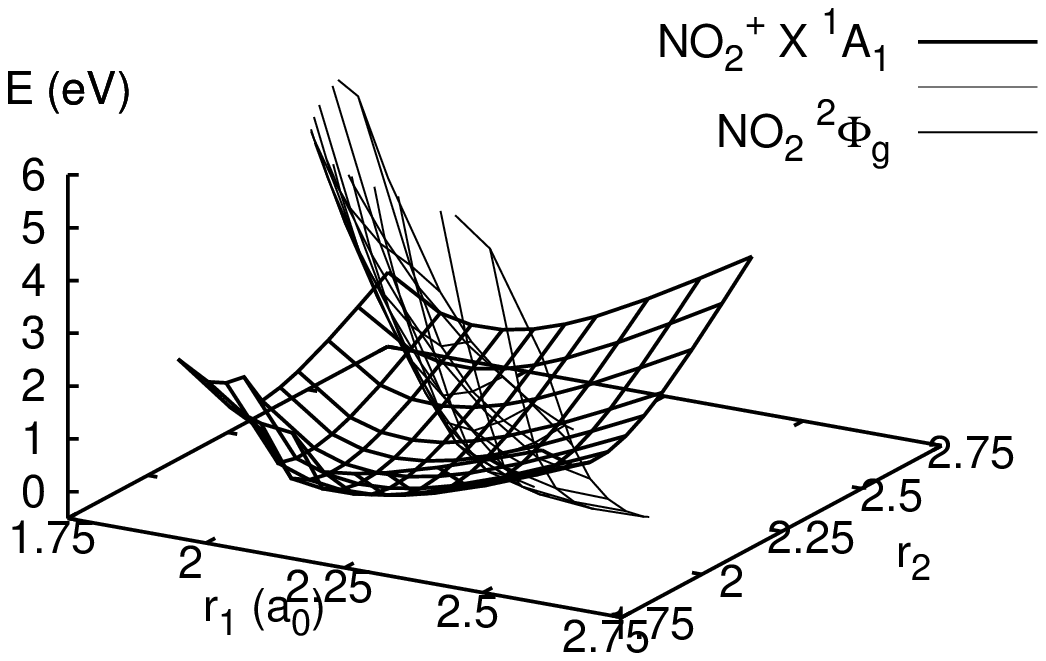}}  \\
\resizebox{0.8\columnwidth}{!}{\includegraphics{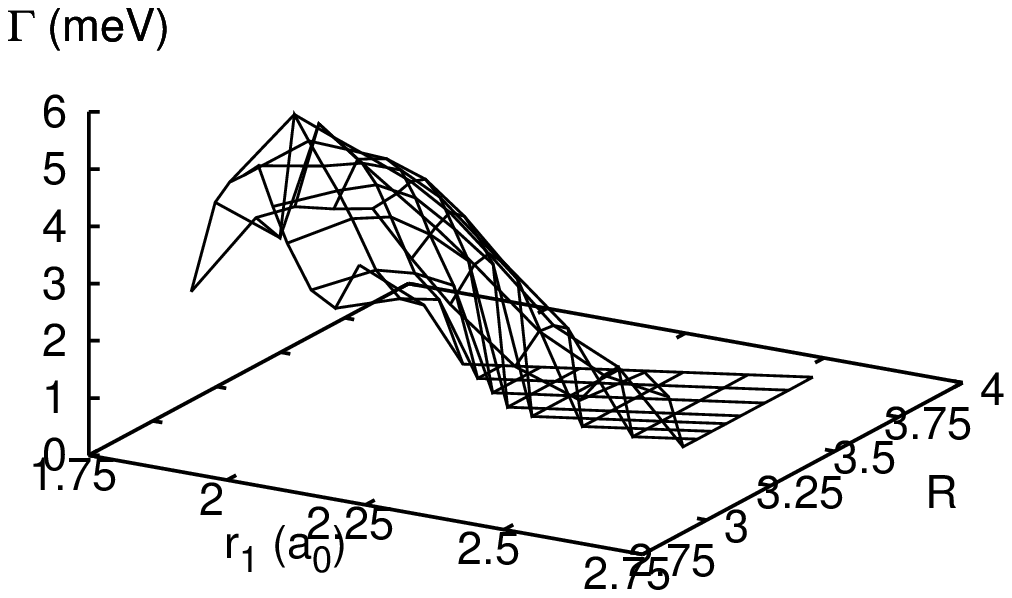}}  \\
\end{center}
\caption{Location (top) and width (bottom) of the 1 $^2$B$_2$ (1 $^2\Phi_g$) Feshbach resonance for linear geometry as a function of bond lengths.}
\label{fig2}
\end{figure}

It is possible estimate the magnitude of the direct DR rate for NO$_2$ + $e^-$ via the components of the 1 $^2\Phi_g$ state, which crosses the cation state closer to its Franck-Condon region.  To accomplish this, three-dimensional complex-valued splined representations of the location and width are developed for 
this $^2$B$_2$ Feshbach resonance.  The location and width of the calculated resonance at linear geometry are shown in Fig.~\ref{fig2}.

The present estimate is based on the standard first order perturbative expression~\cite{omalley} for the DR cross section, with the assumption that the survival 
probability is close to unity here.  We use
the reflection principle~\cite{reflection} approximation to the bound-continuum matrix element,
\begin{equation}
\sigma_{DR}(E)  = \frac{4\pi^2}{k_i^2} \sum_f \vert\langle \Psi_i \vert V_{\bar{a}} \vert \chi_f(E)\rangle \vert^2 
\end{equation}
\begin{equation}
\begin{split}
\displaystyle
\sum_f  & \left\vert \left\langle \Psi_i \left\vert V_{\bar{a}} \right\vert \chi_f(E) \right\rangle \right\vert^2 \\
 & \approx \ \int \ d\vec{R} \ \delta\left(E-\langle \hat{T} \rangle - V(\vec{R}) \right) \left\vert \Psi_i(\vec{R}) V_{\bar{a}} \right\vert^2 
 \end{split}
\end{equation}

The ground vibrational state is approximated as a Gaussian in the normal modes but the asymmetric stretch has been omitted 
despite having calculated three dimensional surfaces.  This omission allows
us to treat the $^2$A$_2$ and $^2$B$_2$ components of the 1 $^2\Phi_g$ state separately.

Fig.~\ref{struct} shows results using a two dimensional vibrational treatment in both the bend and symmetric stretch coordinates, and a one-dimensional treatment including the bend degree of freedom only.  The static degree(s) of freedom are fixed at their equilibrium values.  Evidently, the
cross sections via $^2$B$_2$ are higher; this follows from the larger electronic decay width of this state.

We expect the bending degree of freedom to be most important for the direct DR mechanism, because it is in that degree of freedom that we find
crossings of the neutral and cation states close to the minimum of energy of the Born Oppenheimer potential energy surface.  Indeed, 
the cross sections obtained from the treatment with
only the bending mode are similar to those calculated using two degrees of freedom, validating this hypothesis.
These results should be regarded as approximate because there is considerable uncertainty in the exact location of the cation-neutral crossings.

\begin{figure}
\begin{center}
\resizebox{0.8\columnwidth}{!}{\includegraphics{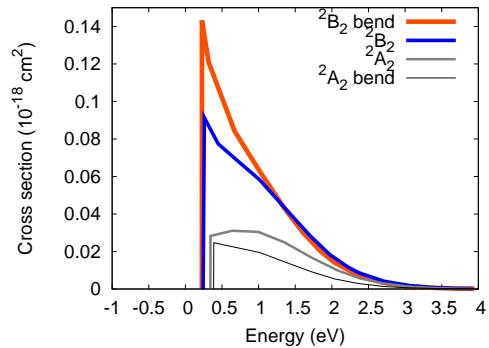}}  \\
\end{center}
\caption{(Color online) Direct DR cross section versus incident electron energy calculated for DR via the 1 $^2\Phi_g$ state.}
\end{figure}

\begin{figure}
\begin{center}
\resizebox{0.8\columnwidth}{!}{\includegraphics{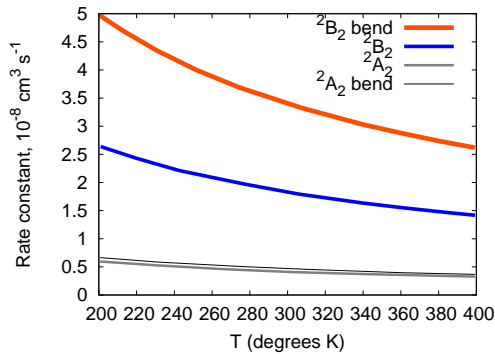}}  \\
\end{center}
\caption{(Color online) Thermally averaged rate coefficient for direct DR via the 1 $^2\Phi_g$ state.}
\end{figure}

\section{Indirect mechanism formulation of direct DR via Rydbergs}

We follow the method originally outlined in Ref.~\cite{hamilton}.  This method uses
Siegert states in place of the bound vibrational states of the usual vibrational 
frame transformation.   These vibrational states $\chi_i$ are not orthogonal with
respect to the usual hermitian inner product.   In the present
case, as in our previous work on the high energy peak of HeH$^+$\cite{haxtonheh}, we include 
cation channels that are energetically open.  An improved formula is used to calculate
the unitary defect $D_i$ of the S-matrix for an electron incident
in channel $i$ , not
\begin{equation}
D_i =  1 - \sum_j \big\vert S_{ij} \big\vert^2
\end{equation}
but
\begin{equation}
D_i = 1 - \sum_{jk} S_{ij} U_{jk} S_{ik}^* 
\end{equation}
\begin{equation}
\qquad U_{jk} = \int \ dR \ \chi_i(R) \chi_j(R)^* \ .
\end{equation}
Following the previous work using our method we define the DR cross section 
as proportional to the unitary defect of the 
frame-transformed S-matrix.

\begin{figure}
\resizebox{0.85\columnwidth}{!}{\includegraphics{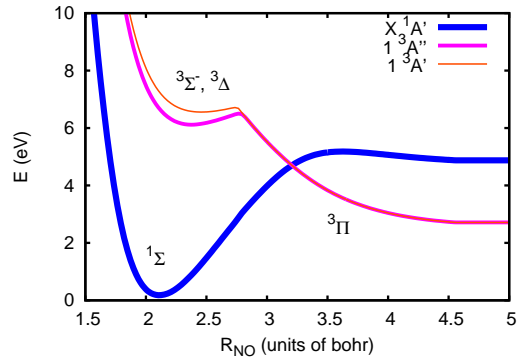}}
\caption{(Color online) Born oppenheimer electronic states of NO$_2^+$ included in the present indirect calculation: energies for $r_2$=2.12$a_0$, linear geometry, with respect to $r_1$.  \label{no2curves}}
\end{figure}

Molecular orbitals are calculated for the cation using the MOLPRO quantum chemistry suite~\cite{MOLPRO}.  
The calculation treats 16 electrons in 11 orbitals, with only the 1$s$ orbitals 
constrained to be doubly occupied, 
and minimize the average energy of the $^1$A$'$, $^3$A$'$, and $^3$A$''$ states with respect to variations
in the orbital and configuration coefficients. 
The first 
is the ground
state at the equilibrium geometry.  The next are the lowest triplet states A' and A'' symmetry.  Within the Franck-Condon
region in d$_{\infty H}$ geometry, these are the 1 $^3\Sigma_u^-$ and one component of the 1
$^3\Delta_u$ state.  These states change
discontinuously where they are crossed by the 1 $^3\Pi_g$ state, each crossing being an actual intersection due to symmetry.
The potential 
energy curves of these states are plotted in Fig~\ref{no2curves}.  We use
these orbitals in another calculation using the UK R-matrix codes, including these three target states
in the scattering calculation.  We obtain the $R$-matrix from these calculations and solve for the MQDT S-matrix.  
We find that the calculation is very well converged with the three target states and a sizable
virtual orbital basis included for penetrating configurations.   We include $s$, $p$, $d$, and $f$-wave scattering orbitals and obtain a multichannel S-matrix with all such partial waves for each of the
three included target states.

\begin{figure}
\resizebox{0.85\columnwidth}{!}{\includegraphics{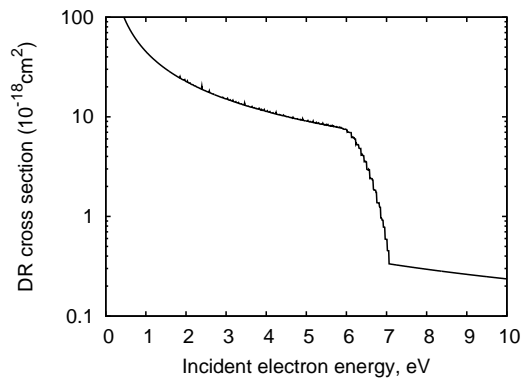}}
\caption{Total indirect (direct via Rydbergs) probability from 1D frame transformation calculation using energy- and R-independent quantum defects.  \label{indirectfig} }
\end{figure}

This treatment is based on the three electronic state MQDT S-matrix calculated at zero energy in the 1 $^1$A$'$ channel (such that the triplet channels are closed) at the equilibrium geometry of the cation.  A very simple treatment is adopted in which the $R$-dependence of the quantum defect is neglected 
(and the energy dependence is likewise neglected).  Such a treatment was sufficient to yield a sensible calculation of the the high energy
DR peak of HeH$^+$~\cite{haxtonheh}.  We include only the motion of single bond stretch at linear geometry.  The results are shown in figure~\ref{indirectfig}.

\section{\textbf{Conclusion}}

We have presented the first estimate of dissociative recombination rates of NO$_2^+$.  Several results analyzing the contribution of direct and indirect processes to dissociative recombination of NO$_2^+$ + $e^-$ were presented.  
These results compared the relative contribution of the direct and indirect mechanisms that drive dissociative recombination.  These calculations 
in reduced dimensionality indicate that the direct DR mechanism is likely to dominate the indirect, the former of which is driven by intersections of neutral 
Born-Oppenheimer potential energy curves with that of the cation.  We find that the 1 $^2\Phi_g$ and the 1 $^2\Pi_g$ electronic states are good candidates 
for states that drive the direct mechanism.  There is considerable uncertainty in the present results, both due to the difficulty in locating the aforementioned 
crossings, and due to the reduced dimensionality of the calculations.  However, these results provide guidance for experimental studies of dissociative 
recombination in these molecules, and also provide information that is useful in the description of the decay channels of Rydberg states of the molecule, 
a topic that has received significant interest in the literature.

\section{\textbf{Acknowledgments}}

This work was supported by the United States Department of Energy Office of Science.

\bibliography{no2.bib}

\end{document}